\documentstyle[12pt]{article}

\begin{document}
\begin{center}
{\bf 
Probing deviation of tribimaximal mixing and reach 
of $\theta_{13}$ using neutrino factory at CERN and 
 ICAL detector at INO}
\end{center}
\vskip 3mm
\begin{center}
{\bf Debasish Majumdar, Ambar Ghosal, Sudeb Bhattacharya, Kamales Kar }
\end{center}
\begin{center}
{\it Saha Institute of Nuclear Physics,
1/AF Bidhannagar, Kolkata 700 064, India
}
\end{center}
\vskip 3mm
{\small
We investigate the deviation from tribimaximal mixing value and the reach of $\theta_{13}$ 
using neutrino factory at CERN and ICAL detector at INO. 
}

\vskip 4mm 
\noindent  
High energy neutrino beams are now playing a very crucial role in determining
the yet unknown features of neutrino physics. Determination of the
neutrino mass hierarchy and determination of the mixing angle 
$\theta_{13}$, probing CP violation effects
at neutrino sector are the rich physics possibilities envisaged 
from the neutrino beam of a neutrino factory. 

In the present work, we explore the possibility 
to observe the deviation, if any, of the neutrino mixing angles 
from their tribimaximal mixing \cite{tri-bi} values 
using a long baseline neutrino experiment 
with the neutrino beam from muon storage ring at CERN and the detector at the
India-based Neutrino 
observatory (INO) covering the distance of about 7152 Km. 
The tribimaximal mixing values for neutrino oscillations are given by
$\sin\theta_{12} = 1/\sqrt{3} = \sin\theta_{\odot}$, 
$\sin\theta_{23} = 1/\sqrt{2} = \sin\theta_{\rm atm}$ 
and $\theta_{13} = 0^o$ \cite{Ma}.
These results exactly follow from the theory with $A4$ group symmetry 
\cite{Ma}. 
It is also to be noted that the CERN-INO baseline length is close to 
the ``magic baseline" 
length -- the baseline length at which the CP effects on neutrino oscillation 
become virtually insignificant and hence reduce greatly the possibility  
of confusing matter effects on neutrino oscillation with the CP violation 
effects.  

The detector at INO site is a magnetized Iron Calorimeter (ICAL) 
\cite{ino}
which has the unique ability to determine the sign of the charge 
of the particle 
passing through it. Thus it has the capability to distinguish between the 
signature of $\bar{\nu_e} \rightarrow \bar{\nu_\mu}$ oscillation 
(appearence channel, $\mu^+$ signal at ICAL) and the survival of $\nu_\mu$ 
($\nu_\mu \rightarrow \nu_\mu$ oscillation in disappearence
channel; $\mu^-$ signal at ICAL) using the neutrino beam. 
The calculations done here are for the ICAL detector of is 50 kTon 
mass and with an exposure time of 5 years. 


The reach of $\theta_{13}$ for ICAL at INO detector is also investigated. 
This is defined as the lowest value of $\theta_{13}$ that can be probed 
by ICAL for different energies of initial decaying muons at the 
horn of the neutrino factory.

Neutrino beams are generated from the decay of muons, 
$\mu^{\pm}\rightarrow e^{\pm}+\nu_e(\bar{\nu_e})+\nu_\mu(\bar{\nu_\mu})$
from the srtaight section of the muon storage ring 
\cite{ref2,ref3}. For unpolarized muon beam, 
the flux distributions of $\nu_\mu$ and $\bar\nu_e$ are  given by 
$$
{\left({\frac{d^3N}{dtdAdE_\nu}}\right)}^{\nu_\mu}_{lab} 
= 
\frac{4g_{lab}J^2}{\pi L^2 E_\mu^3}E_{\nu_\mu}^2\left(3 - 4\frac{E_{\nu_\mu}}
{E_\mu}\right)
\eqno(1)
$$  
$$
{\left({\frac{d^3N}{dtdAdE_\nu}}\right)}^{\bar\nu_e}_{lab} 
= 
\frac{28g_{lab}J^2}{\pi L^2 E_\mu^4}E_{\bar\nu_e}^2\left(E_\mu 
 - 2E_{\bar\nu_e}\right)
\eqno(2)
$$  
where $g_{lab}$ is the number of muons produced, 
$J$ is the Jacobian factor arising due to transformation 
from rest frame to lab frame and is given by 
$$
J = \frac{1}{\gamma(1-\beta\cos\theta)}
\eqno(3)
$$
$E_\mu$ is the energy of the decaying muon and  
$E_{\nu_\mu}$ and $E_{\bar\nu_e}$ are energies of the 
neutrinos produced due to decay of these muons. Also, $\gamma$ is 
the boost factor and 
$\beta = p_\mu/E_\mu$. The angle $\theta$ is the off axis angle which 
we set to zero. 

For our analysis, we set the parameters as 
$g_{lab} = 0.35\times {10}^{20}$ considering $35\%$ efficiency 
of the produced muon number. 
The parameter $L$ is the length traversed by the neutrino 
from the source to the detector through the earth. In the 
present work, as mentioned, we take $L = 7152$ Km which is the 
distance 
between the source at CERN to the detector at the INO site 
at PUSHEP ($11^o5^`$N, $76^o6^`$E) \cite{sanjib}.  
The distribution of flux for both $\bar\nu_e$ and $\nu_\mu$ 
are shown in Fig. 1 and Fig. 2 respectively, 
where $E_{\bar\nu_e}$ varies from 0-10 GeV and 
$E_{\nu_\mu}$ varies from 0-15 GeV and we observe for both the cases 
a large 
number of muons hitting the detector.

We consider here a three flavour neutrino oscillation with matter 
effects \cite{huber} due to the passage of the neutrino through the earth. 
It may be noted here that the average earth matter density for 
CERN-INO baseline can be calculated to be 4.15 gm/cm$^3$.   
\vskip 0.1in
\noindent
We consider the $\nu_\mu\rightarrow\nu_\mu$ as well as 
$\bar\nu_e\rightarrow\bar\nu_\mu$ mode in the three flavour 
oscillation scenario. The interaction 
of $\nu$ with the ICAL (Iron Calorimeter) detector proposed
at INO site are mainly arising through quasi-ealstic (QE) and 
Deep inealstic scattering (DIS). At low energy we have considered 
QE mode ($E_\nu \leq 1$GeV), and for $E_\nu >1$GeV the dominant 
contribution will come from DIS.

We emphasise here that the resolution function 
of the detector obtained 
from the exact simulation of the ICAL detector using GEANT 3.2 simulation
code has been included. 

We have calculated the total muon yields for CERN-INO baseline  
for the deviation from tribimaximal mixing and the maximum and the minimum 
muon yield values
for such deviation are tabulated in Table 1. In order to estimate 
the deviation, we take the experimentally obtained range for 
the three mixing angles given as $30^o \leq \theta_{12} \leq 38^o$,
$38^o \leq \theta_{23} \leq 53^o$ and $\theta_{13} < 12^o$.  
The calculation is 
repeated for several values of the CP violation phases, $\delta$. 

The wrong sign muon yield for the best fit values of the oscillation 
parameters \cite{huber} is also calculated for the CERN-INO distance. 
This is tabulated in Table 2. 

\begin{center}
\begin{tabular}{|c|c|c|c|c|c|}
\hline
CERN-INO & $\delta$ & $\theta_{13}$ & \multicolumn{2}{c|} {Muon yield}& Boost \\
\cline{4-5}
&CP&&Max & Min & \\
\hline
&&&&& \\
50 kton  & 0$^o$        & 9$^o$             & 605835 & 486304 & \\
5 years  & 0$^o$        & 6$^o$             & 613035 & 483768 & 472.0 \\
         & 270$^o$      & 6$^o$             & 626210 & 491390 & $E_\mu$=50 GeV \\
\hline
\end{tabular}
\end{center}
\noindent {\bf Table 1.} {\small Variations of muon yield at INO with 
the deviation from tribimaximal mixing for 
neutrino beam from a neutrino factory at CERN} 

\begin{center}
\begin{tabular}{|c|c|c|c|c|c|}
\hline
CERN-INO & $E_\mu$ & \multicolumn{2}{c|}{Muon Yield} & $\delta$ & $\theta_{13}$ \\
\cline{3-4}
& & Wrong  & Right  & &  \\
         & (GeV)   &  sign $\mu$  &  sign $\mu$  & CP       &   (in degrees) \\
\hline
&&&&& \\
50 kton  & 50        & 79447             & 458893 & 0$^o$ &  6\\
5 years  &           & 85834             & 449788 & 0$^o$ & 9 \\
         &           & 72827           & 477637 & 90$^o$ & 6 \\
         &           & 75857           & 473150  & 90$^o$ & 9 \\
\hline 
\end{tabular}
\end{center}
\noindent {\bf Table 2.} {\small The wrong sign muon yield at 
INO for a neutrino beam from CERN.}
\vskip 0.1in

The reach of $\theta_{13}$ at the ICAL detector at INO is also addressed.
For this purpose, we first calculate the wrong sign muon yield 
for $\theta_{13} =  0.0$. Then we investigate the minimum value 
of $\theta_{13}$ that will be required to obtain the wrong sign 
muon yield with 1$\sigma$ deviation. In this case we consider 
the statistical error only.
The process is repeated for the inverted 
mas hierarchy as well.

In Fig. 3, the $\sin^2 2\theta_{13}$ reach,
thus calculated, is shown for different values of muon energies 
(i.e. for different values of the boost) that are allowed to decay in a 
neutrino factory. Here too the CERN-INO baseline is considered. 
There is no significant difference for the $\theta_{13}$ reach
values for normal and inverted mass hierarchies. 
It is also seen from Fig. 3 that INO is capable of detecting 
$\theta_{13}$ as low as $\sim 1^o$ for $E_\mu \sim 25$ GeV. The 
reach is further reduced to $\sim 0.2^o$ for higher boost (higher 
$E_\mu \sim 60$ GeV). This is an encouraging result. 

A detailed Geant-based study is under progress.

The authors thank S. Goswami and R. Gandhi for helpful discussions.


\newpage
\begin{center}
{\bf Figure Captions}
\end{center}

\noindent {\bf Fig. 1} The $\bar {\nu_e}$ flux from a neutrino
factory with muon energy $E_\mu = 50$ GeV. See text
for details.

\vskip 2mm

\noindent {\bf Fig. 2} The $\nu_\mu$ flux from a neutrino
factory with muon energy $E_\mu = 50$ GeV. See text
for details.

\vskip 2mm

\noindent {\bf Fig. 3} 
The reach of $\sin^2 2\theta_{13}$ at INO
for neutrinos from a neutrino factory at CERN for various enrgies $E_\mu$
of decaying muons. See text for details.

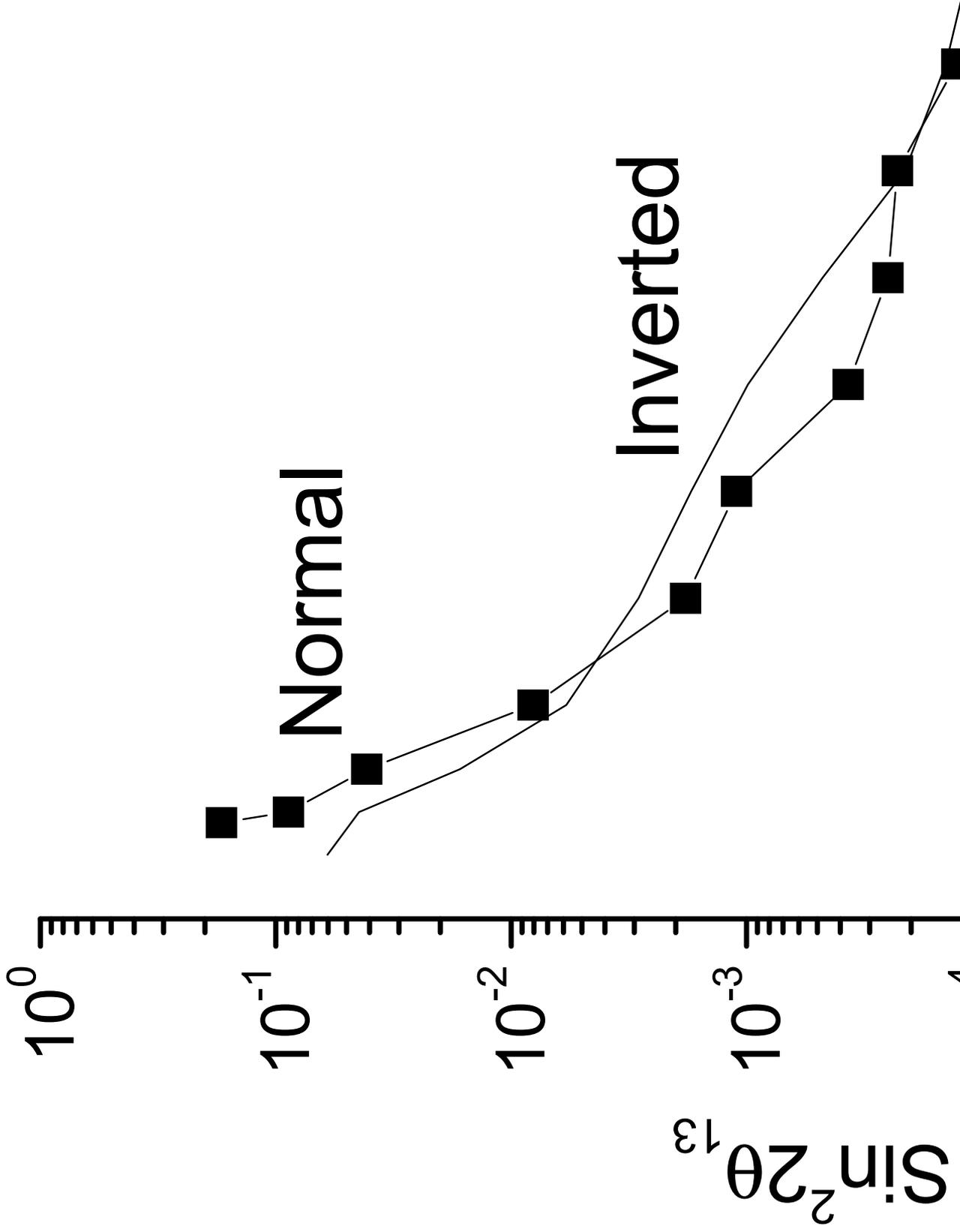
\end{document}